\documentclass[journal]{IEEEtran}
\usepackage{blindtext}
\usepackage{graphicx}
\usepackage{amsmath,amssymb}
\usepackage{lineno}
\usepackage{color}
\usepackage{subfigure}
\usepackage{multirow}
\usepackage{algorithm}
\usepackage{fancyhdr}
\usepackage{cases}
\usepackage{epsfig}
\usepackage{epstopdf}
\usepackage{array,color}
\usepackage{amsmath}
\usepackage{stfloats}
\usepackage{url}
\usepackage{algorithm}
\usepackage{algorithmic}
\usepackage{amsfonts,amssymb}
\usepackage{amsmath}
\usepackage{bm}
\usepackage{booktabs}
\usepackage{array}
\usepackage{mathrsfs}
\usepackage{graphicx}
\usepackage{cite}
\usepackage{stfloats}
\usepackage{float}
\usepackage{eqnarray}
\usepackage{hyperref}
\usepackage{makecell}
\hypersetup{hidelinks}
\usepackage{booktabs}

\DeclareUnicodeCharacter{2061}{X}
\begin{document}
\begin{sloppypar}
\title{
Deep Learning-based Eye-Tracking Analysis for Diagnosis of Alzheimer's Disease Using 3D Comprehensive Visual Stimuli}

\author{Fangyu~Zuo,
        Peiguang~Jing,
        Jinglin~Sun,
        Jizhong Duan,
        Yong~Ji,
        Yu~Liu,~\IEEEmembership{Member,~IEEE,} 
\thanks{Fangyu Zuo, Jinglin Sun and Yu Liu are with the School of Microelectronics, Peiguang Jing is with School of Electrical and Information Engineering, Tianjin University, Tianjin, China. 
Jizhong Duan is with the Faculty of Information Engineering and Automation, Kunming University of Science and Technology, Kunming, China.
Yong~Ji is with Tianjin Key Laboratory of Cerebrovascular and Neurodegenerative Diseases, Department of Neurology, Tianjin Dementia Institute, Tianjin Huanhu Hospital, Tianjin, China.
\textit{Corresponding author: Yu Liu, liuyu@tju.edu.cn.}} }
\markboth{A SUBMISSION TO ***}%
{Shell \MakeLowercase{\textit{et al.}}: Bare Demo of IEEEtran.cls for Computer Society Journals}
\IEEEcompsoctitleabstractindextext{
\begin{abstract}
 Alzheimer's Disease (AD) causes a continuous decline in memory, thinking, and judgment. Traditional diagnoses are usually based on clinical experience, which is limited by some realistic factors. In this paper, we focus on exploiting deep learning techniques to diagnose AD based on eye-tracking behaviors. Visual attention, as typical eye-tracking behavior, is of great clinical value to detect cognitive abnormalities in AD patients. To better analyze the differences in visual attention between AD patients and normals, we first conduct a 3D comprehensive visual task on a non-invasive eye-tracking system to collect visual attention heatmaps. We then propose a multi-layered comparison convolution neural network (MC-CNN) to distinguish the visual attention differences between AD patients and normals. In MC-CNN, the multi-layered representations of heatmaps are obtained by hierarchical convolution to better encode eye-movement behaviors, which are further integrated into a distance vector to benefit the comprehensive visual task. Extensive experimental results on the collected dataset demonstrate that MC-CNN achieves consistent validity in classifying AD patients and normals with eye-tracking data.
\end{abstract}
}
\maketitle
\IEEEdisplaynotcompsoctitleabstractindextext
\IEEEpeerreviewmaketitle
\section{Introduction}
\IEEEPARstart
{R}{ecently}, the prevalence of Alzheimer’s Disease (AD) has been increasing rapidly, with an estimated 106.8 million worldwide cases by 2050 \cite{brookmeyer2007forecasting}. AD is an irreversible and chronic neurodegenerative brain disorder characterized by abnormal cognitive symptoms, including speech and language impairments, and memory declination \cite{mckhann1984clinical}. As the most common cause of dementia, AD is estimated to be the fifth leading cause of death in elderly people. To date, treatments for AD only delay the disease onset or slow its progression. Thus, it is crucial to diagnose AD in the early stages by suitable biomarkers for a more aggressive therapy to overcome the symptoms and prevent its deterioration to dementia \cite{readman2021}. Currently, the clinical diagnoses of AD  are usually made by experienced clinicians through neuropsychological tests \cite{folstein1983mini}, blood detections \cite{olsson2016csf}, and medical images \cite{patro2019early}. Although these methods have become mainstream diagnostic methods for AD, several limitations still hinder their wider application. For example, neuropsychological tests generally need to take 10–15 min to complete under the guidance of professionally trained medical staff. This will cause a significant consumption both of time and manpower. Moreover, the collection of blood samples is invasive and will cause harm of some degree to the patients. Medical imaging requires sophisticated instrumentations and manual analysis that is relatively inefficient. Therefore, finding new biomarkers that can be noninvasively detected in the early stage of AD as well as be analyzed by artificial intelligence techniques like deep learning is highly desirable. To alleviate these issues, we propose a novel deep-learning-based approach to diagnose AD using eye-movement data.

As biomarkers that can be detected in the early stage of AD, eye movements have shown great potential for AD diagnosis, which is demonstrated in extensive literature \cite{anderson2013,biondi2017eye,hedayatjoo2023comparison,coors2022associations}. Since eye movements are easy to be collected by eye-tracking tools, a growing body of research provides evidence that eye-tracking is a simple and non-invasive method for AD diagnosis. For example, numerous studies have shown that abnormal saccades are associated with AD \cite{garbutt2008oculomotor,kahana2018prosaccade,kaufman2012executive}. Various eye-movement tasks have been designed to capture the abnormal saccades in AD patients, which include prosaccade tasks that require participants’ eyeballs to move forward to a visual stimulus, and antisaccade tasks that require subjects’ eyeballs to move away from the presented stimulus \cite{opwonya2022saccadic}. It is noted that AD patients usually have longer delays and decreased saccade amplitudes in initiating saccades compared with normals \cite{garbutt2008oculomotor}. In addition, since cognitive impairment in AD patients is often accompanied by visual attention deficits, attention abnormalities have been well characterized in detecting AD. Previous researches also find that eye-movement behavior reveals multiple visual attention processes, including emotional attention \cite{bourgin2018early}, visual exploration strategies \cite{maruff1995, foldi1992, nebes1989}, and color preferences \cite{wang2015, stanzani2019color}.

To explore the visual attention behaviors in AD patients, several visual tasks with eye-tracking have been designed in current studies, most of which require participants to respond to instructions verbally or behaviorally \cite{boucart2014animal,kawagoe2017face,mosimann2004visual}. For example, Kawagoe et al. \cite{kawagoe2017face} designed a visual memory task on 18 patients with mild cognitive impairment (MCI) which is a prodromal stage of Alzheimer’s disease, and 18 normals. Participants were asked to view a study stimulus for 3s followed by a 3-5s gap. Then a test stimulus pair was presented to them. Participants were required to respond whether either (left or right) or neither test stimulus was the same as the study stimulus by pressing one of three buttons. Mosimann et al. \cite{mosimann2004visual} conducted a clock reading task on 24 AD patients and 24 normals by presenting participants with clocks of different times and asking them to read and state the times. They decided whether to read the next clock by pressing the mouse buttons. In these tasks, results mainly focused on two indices for eye-movement data: fixation duration and the number of fixations separately in the divided Areas of Interest (AOI). By analyzing the experimental results, the researchers found that patients show different attentions to the same stimulus. For instance, patients focused less on the mouth area of a face during the memory test. In the clock reading test, patients paid less attention to the clock hand areas. As indicated in the review literature, such tasks combined with eye-tracking technology show particularly promising results for the distinction between AD patients and normals. However, instructions will easily cause the subjects to be nervous and uneasy, resulting in less objective evaluation scores. Daffner et al. \cite{daffner1992} conducted a viewing task without instructions by presenting participants with well-designed slides, and found that AD patients focus less on novel elements in a photograph.

Deep-learning-based models have been shown to play a crucial role in identifying AD patients with high sensitivity \cite{tufail2020binary,lee2019predicting,higdon2004comparison}. These models are commonly used to deal with medical imaging such as positron emission computed tomography (PET) or magnetic resonance imaging (MRI), mainly because the feature representations produced by these models can be helpful even if the data is partially missing. In addition to medical images, deep-learning models combined with eye-tracking technology have also shown good performance in detecting cognitive impairments in AD patients \cite{przybyszewski2023machine,yin2023internet}. Biondi et al. \cite{biondi2017eye} built a deep neural network using the trained autoencoders and a Softmax classifier that allows identifying AD patients with 89.78$\%$ of accuracy based on eye-tracking data. However, deep-learning models for AD recognition or classification based on eye-movement data are rare in existing studies due to the lack of large-scale eye-tracking datasets. Sun et al. \cite{sun2022novel} proposed a nested autoencoder model to identify AD patients and normalattentioneye-movement dataset on the designed visual memory task, which showed 85$\%$ average accuracy in AD recognition.

The above encouraging results indicate that deep-learning-based models are promising for understanding the dynamics of eye-movement behavior and its relationship to underlying cognitive impairments of AD patients. However, several limitations have hindered the application of such researches. For example, due to the insufficiency of existing eye-tracking system functions and algorithms in 3D displays, most of the current studies concentrate on visual stimulus of 2D display, and few studies have used a 3D display form. Even though it has been pointed out that under the same content, 3D stimuli can stimulate more abundant eye movements and brain activity behaviors than 2D stimuli \cite{arora2021deep}. Moreover, further research is limited by a lack of the large-scale eye-tracking dataset since the performance of deep-learning-based approaches depends heavily on the training dataset’s size.

Motivated by the above limitations, we propose a multi-layered comparison convolution neural network (MC-CNN) to differentiate AD patients from normals with an eye-movement dataset on a designed 3D comprehensive visual task. The hypothesis is that using deep learning to identify the key characteristics of eye behavior during the 3D comprehensive visual task may lead to an accurate classification between AD patients and normals, which can provide meaningful information on the cognitive declination of the patients for clinicians. Our main contributions are summarized as follows:
\begin{itemize}
\item We explore a novel approach to detect the visual attention deficits of AD patients by analyzing eye-tracking data recorded in a 3D comprehensive visual task, by which eye-movement data is promising to be used as a biomarker for early AD diagnosis;
\item A deep-learning-based MC-CNN model with a multi-layered feature extractor is proposed to capture the visual attention features efficiently. Using the integration of feature representations on different features and global average pooling (GAP), the proposed model can transfer the diagnosis of AD to an automatical classification problem via artificial intelligence;
\item By comparing eye-movement heatmap pairs of subjects, an augmented dataset is constructed based on the collected eye-tracking data. The similarity information of visual attention features between individuals is learned by MC-CNN for better classification performance. 
\end{itemize} 
The rest of this paper is organized as follows. Section II introduces the overall process of the experiment and its mathematical principles. Section III reports the experimental evaluation and analysis of our proposed method. Section IV gives the conclusion and future work.
\section{Materials and Methods}
In this section, we present the materials and methods used in the study, including participants, eye-movement data collection and preprocessing, dataset augmentation, the framework of MC-CNN, classification models for comparison, and evaluation metrics.
\subsection{Participants}
Eye-tracking data were obtained from a total of 106 participants including 68 normals and 38 AD patients. The AD patients (23 female, 15 male) were 54$-$80 years old (68$\pm$7 years), and were recruited from the cognitive impairment clinics, Tianjin Huanhu Hospital, Tianjin, China (\href{http://www.tnsi.org/}{http://www.tnsi.org/}). The diagnoses were based on NINCDS-ADRDA and included patient history, clinical impression, and brain pathology as detected by mini-mental state evaluation and structural imaging. In AD patients, humans with uncorrected dysfunctions of vision or hearing loss, mental disorders, or other symptoms that made them unable to complete the proposed visual task or scale assessment were excluded. The normals were recruited from friends and relatives of patients who had no subjective or informant-based complaints of cognitive decline. A subgroup of normals was selected for comparison with AD patients and other normals. All eye-tracking data were presented in the form of heatmaps. The use of eye-tracking data for analysis of AD patients' cognitive decline was approved by the World Medical Association Declaration of Helsinki \cite{world2001world}.
\subsection{Eye-Movement data collection and preprocessing}
We performed a preliminary analysis of eye-tracking data under the 3D comprehensive visual task to compare characteristic differences in visual attention between AD patients and normals, and then determined whether these differences could serve as a detecting tool to identify AD.
\subsubsection{Eye-tracking system with 3D video stimuli}
Eye movements were recorded using a non-invasive eye-tracking system with stereo stimuli designed by Sun et al. \cite{sun2021novel}, School of Microelectronics, Tianjin University, Tianjin, China. This system estimates gaze positions at an average error of 1.85cm/0.15m over the workspace volume 2.4m$\times$4.0m$\times$7.9m. It displays stereo stimuli in the resolution of 1920$\times$1080 pixels in a limited vision without requiring the viewer to wear any accessories, which gives a friendly and immersive 3D visual experience. The system collects eye-movement data binocularly with the eye-tracking module that can be adjusted in a 360-degree direction freely to meet the need of different users. A calibration procedure is applied to all participants. The structure of the system and the demonstration of the eye-movement data collection process are presented in Fig. 1.

\begin{figure}[htbp]
\centering
\includegraphics[width=3.5in]{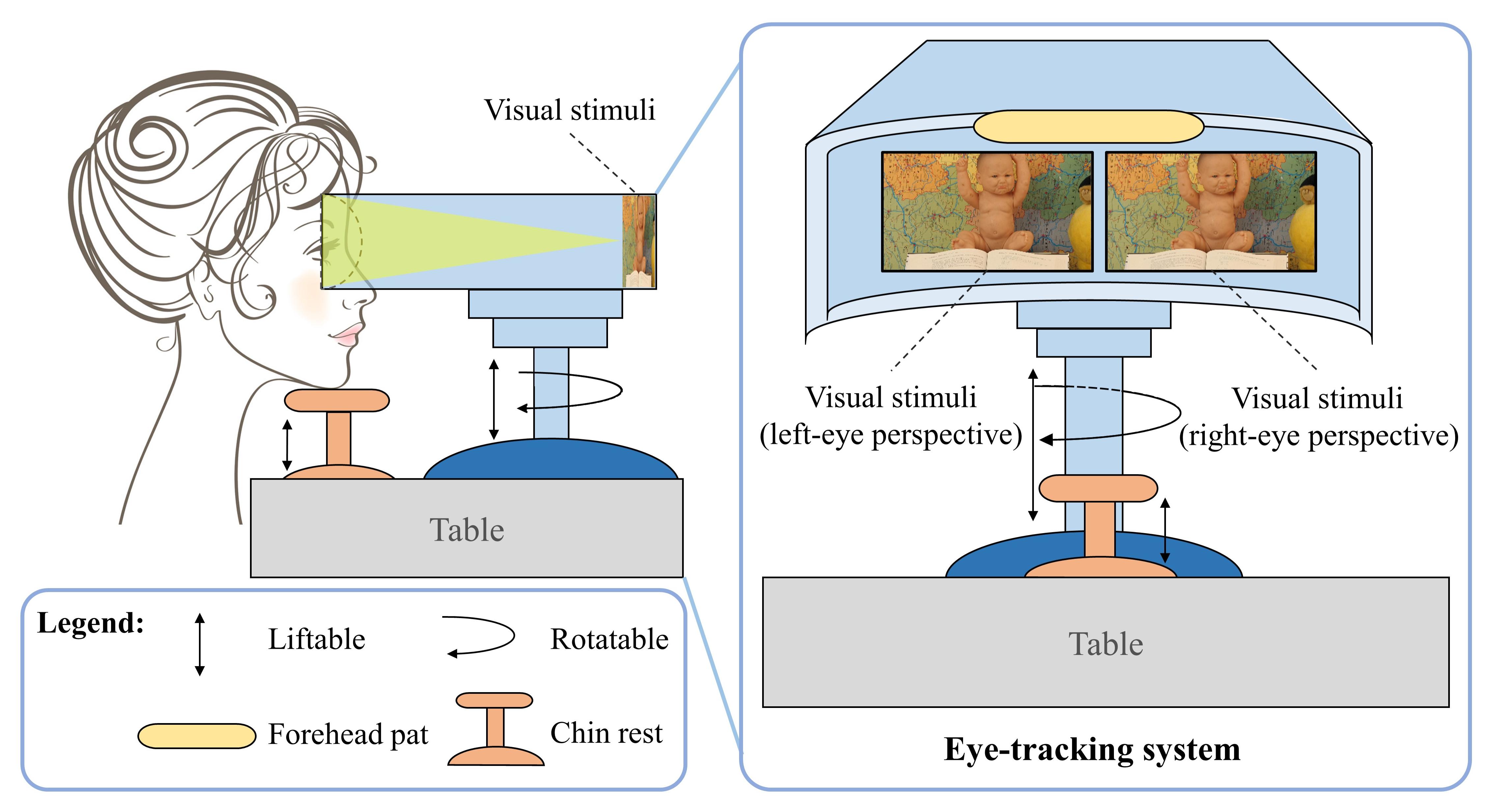}
\caption{The structure of the non-invasive eye-tracking system with stereo stimuli and the eye-movement data collection process based on the system.}
\end{figure}

\subsubsection{3D comprehensive visual task}
The 3D comprehensive visual task was applied based on the above-mentioned eye-tracking system. During the task, participants were seated in a chair in front of the device. As illustrated in Fig. 1, they had to press their heads on the forehead and chin rest in order to minimize head movements. A repeated calibration procedure was used before each personal experimental block to track the gaze accurately. The calibration stimulus was a grid containing nine red dots at random locations that were displayed one at a time on a black background. During calibration, the participants were instructed to fixate on the dot on the screen and to move their eyes to the next dot. Using these reference points, the system creates a mapping function that relates all eye positions to points in the calibration area. Each mark was displayed for 3s to ensure stability. Eye movement recordings were conducted in a quiet room in order to keep viewers in a natural state. Following the calibration, participants performed the 3D comprehensive visual task. They were presented with a series of 3D images for a duration of 5s each with a 1s black background between every two photographs as shown in Fig. 2. A total of 9 images were for testing, which contained different scenes, including natural scenes, human bodies, cartoon characters, etc. Participants were not given any instructions but were only required to freely view the images. It takes approximately 2 minutes on average to complete the entire visual task.   

\subsubsection{Heatmaps generation}
To intuitively represent the eye-movement data, a fixation map will be constructed by Matlab. All measured fixation points for each image are overlapped into a map. Then, the generated fixation map can be smoothed and normalized with a Gaussian kernel to generate a color-coded visualization, i.e., heatmaps. Every participant gets 9 heatmaps corresponding to the 9 photographs presented. These nine heatmaps are then stacked to form a final heatmap for the participant. All the heatmaps will be used as the source input for the MC-CNN to extract discriminative represent features for classifying AD patients and normals. The overall procedure of eye-movement data collection and heatmaps generation is presented in Fig. 2. 

\begin{figure}[htbp]
\centering
\includegraphics[width=3in]{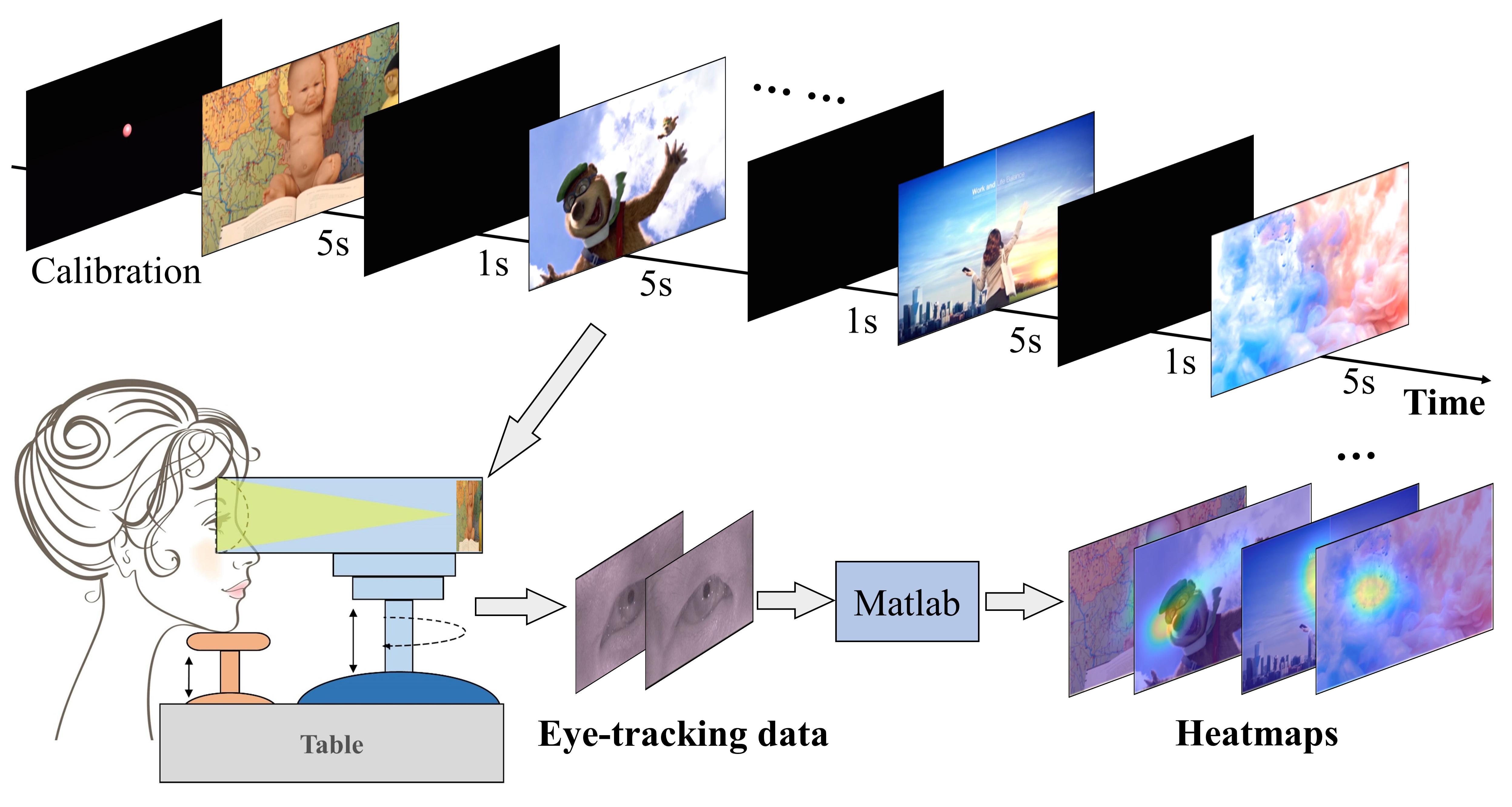}
\caption{The workflow of eye-tracking data collection and heatmaps generation.}
\end{figure}

Since AD patients and normals perform different attentions to the same visual stimuli, their heatmaps show obvious differences. Several sample heatmaps of 3 AD patients and 3 normals that are randomly selected from the participants are illustrated in Fig. 3. As can be observed, visual attention represented by heatmaps of normal participants shows greater similarity, however, the differences between AD patients and normals are apparent. For example, the normal participants are usually good at capturing the main bodies in a picture, especially the human bodies and cartoon characters, while the patient participants pay more attention to the background areas. The preferences of color tunes in a picture are also different between AD patients and normals. 

This particular visual attention phenomenon is further evidence of AD patients’ visual attention deficits. Therefore, it is a reasonable hypothesis that eye-movement data in the comprehensive visual task can reflect the attention processes occurring in the participant’s brain. Eye movements of normals in the task are roughly similar, and eye movements of AD patients show obvious differences. This is of great value in supporting clinical practice in detecting and diagnosing AD through visual attention measured by eye tracking. 

\begin{figure}
    \centering
    \includegraphics[width=3in]{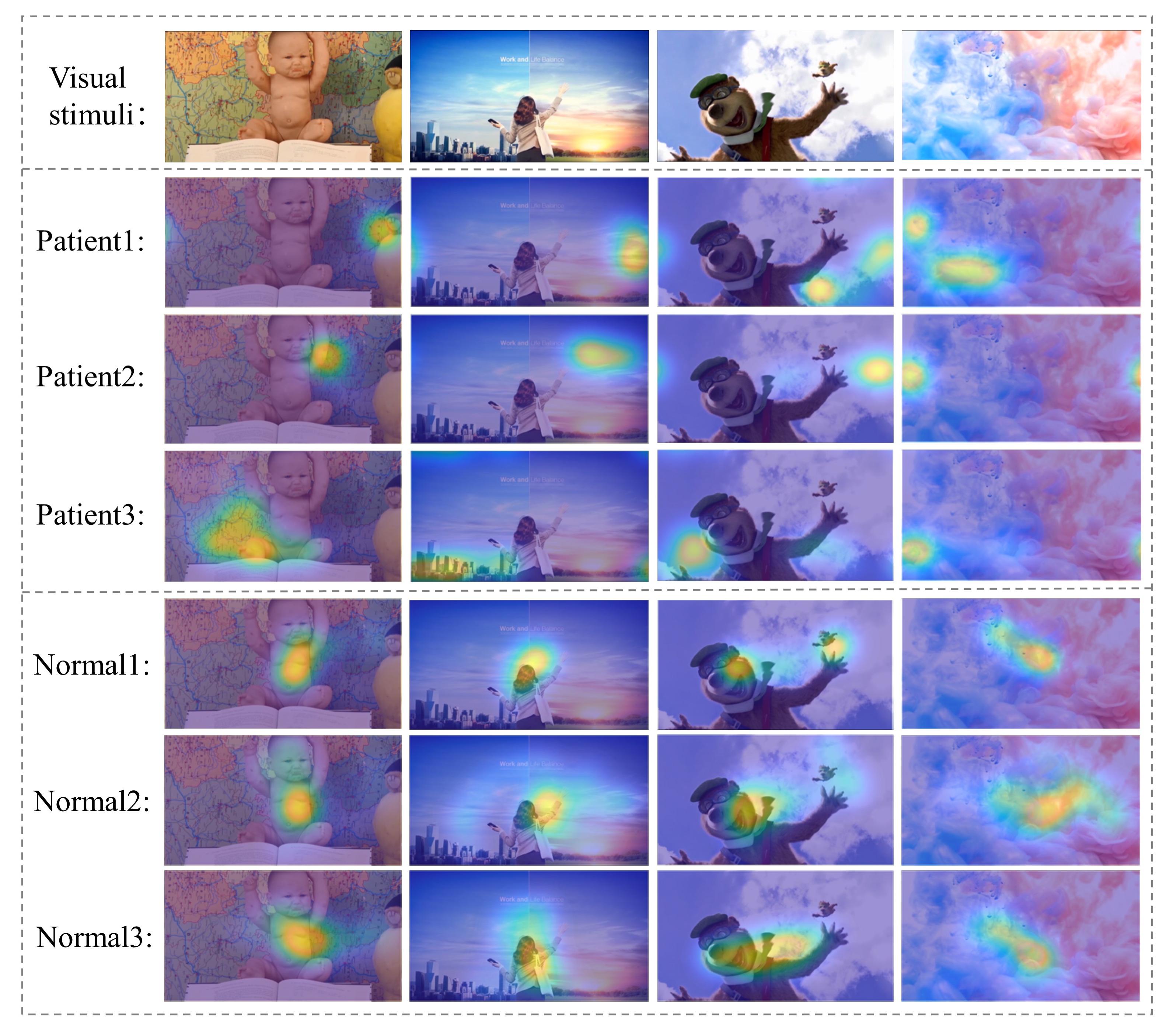}
    \caption{Several sample heatmaps of 3 AD patients and 3 normals that are randomly selected from participants.}
\end{figure}

\subsection{Dataset augmentation}
The training dataset size is essential to deep learning-based approaches. A larger training dataset can provide more details to a deep-learning-based classification model for generalizing patterns of the samples. In this experiment, however, the size of the eye-movement dataset is relatively limited due to patients' privacy as well as organizational challenges. Compared to patients, eye-movement data of normals is easier to be collected, which usually causes an unbalanced distribution of the samples in the dataset. To address the problem presented above, we have augmented the training dataset based on the finite eye-tracking data of the recruited participants, which provides a larger and more balanced training dataset for the model. 

In the 68 normal participants we have recruited, heatmaps of 30 individuals are selected to form a normal group. The remaining heatmaps of 38 AD patients and 38 normals respectively form the AD group and the other normal group. Every subject in the AD group and normal group will combine with a normal subject in the other normal group to form a combination, which expands the size of the training dataset to 2280. By pairwise organizing the heatmaps, the number of combinations is twenty times as much as the original heatmaps. As shown in Fig. 4, the similarities of heatmaps in normal$\&$normal combinations are higher than AD$\&$normal combinations. Therefore, AD$\&$normal combinations are labeled as 0 while normal$\&$normal combinations are labeled as 1. The augmented heatmap dataset is then fed into MC-CNN to explore the discriminative features between the AD$\&$normal combinations and the normal$\&$normal combinations.

\begin{figure}[htbp]
\centering
\includegraphics[width=3in]{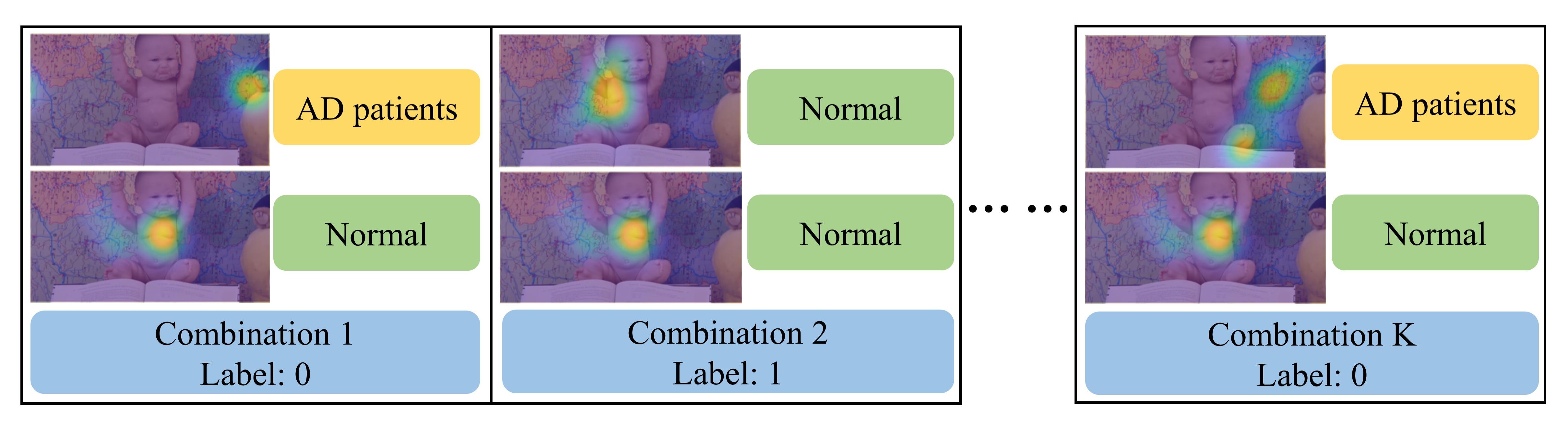}
\caption{The adopted pair-wise sample augmentation scheme.}
\end{figure}

The labeled augmented heatmap dataset is divided into three subsets in the ratio of 6:2:2, namely a training set, a validation set, and a test set, as shown in table I. The training set is learned by MC-CNN, allowing the model to update its parameters with a back propagation procedure, which is called the convergence process. The validation set is used to determine the current converging state of the model, while the test set is used to confirm its final performance.

\begin{table}[htbp]
\caption{Division of the augmented heatmap dataset.}
 \centering
\begin{tabular}{ccccc}
\toprule
Combination &Training & Validation & Test & Label\\
\midrule
AD \& Normal & 660 & 240 & 240 & 0 \\
Normal \& Normal & 660 & 240 & 240 & 1 \\
\textbf{Sum} & \textbf{1320} & \textbf{480} & \textbf{480} & \textbf{0$\slash$1}\\
\bottomrule
 \end{tabular}
\end{table}

\subsection{Framework of MC-CNN}
This article proposes a novel classification mechanism by which visual attention in heatmaps can be classified in an end-to-end manner. As shown in Fig. 5, MC-CNN consists of a convolutional feature extractor and a predictor. After data augmentation (see Section II-C), the heatmaps are fed into MC-CNN. The proposed model learns the similarity information between heatmap pairs for classification. More specifically, the feature extractor aims to capture distinguishing feature representations in heatmaps. Then the feature representations are used to calculate the similarity between the heatmap pairs that are separately from a normal and an unknown class. To utilize more diverse features from shallow to deep convolutional features, the multi-layered feature representation is introduced based on hierarchical residual blocks to form a feature extractor. While the feature extractor extracts feature maps of the input heatmaps, the global averaging pooling layer (GAP) transforms the feature maps into feature vectors. In this manner, the distance of feature vectors extracted from heatmap pairs can represent the similarity information. MC-CNN is designed with the following two improvements: 1) Residual blocks and integration of multi-layered representations are exploited in the feature extractor so that features from low level can be reused, and the vanishing gradient problem can be prevented; 2) GAP is used to generate feature vector without adding parameters thus overfitting is avoided in this layer. 

\begin{figure*}[htbp]
\centering
\includegraphics[width=6in]{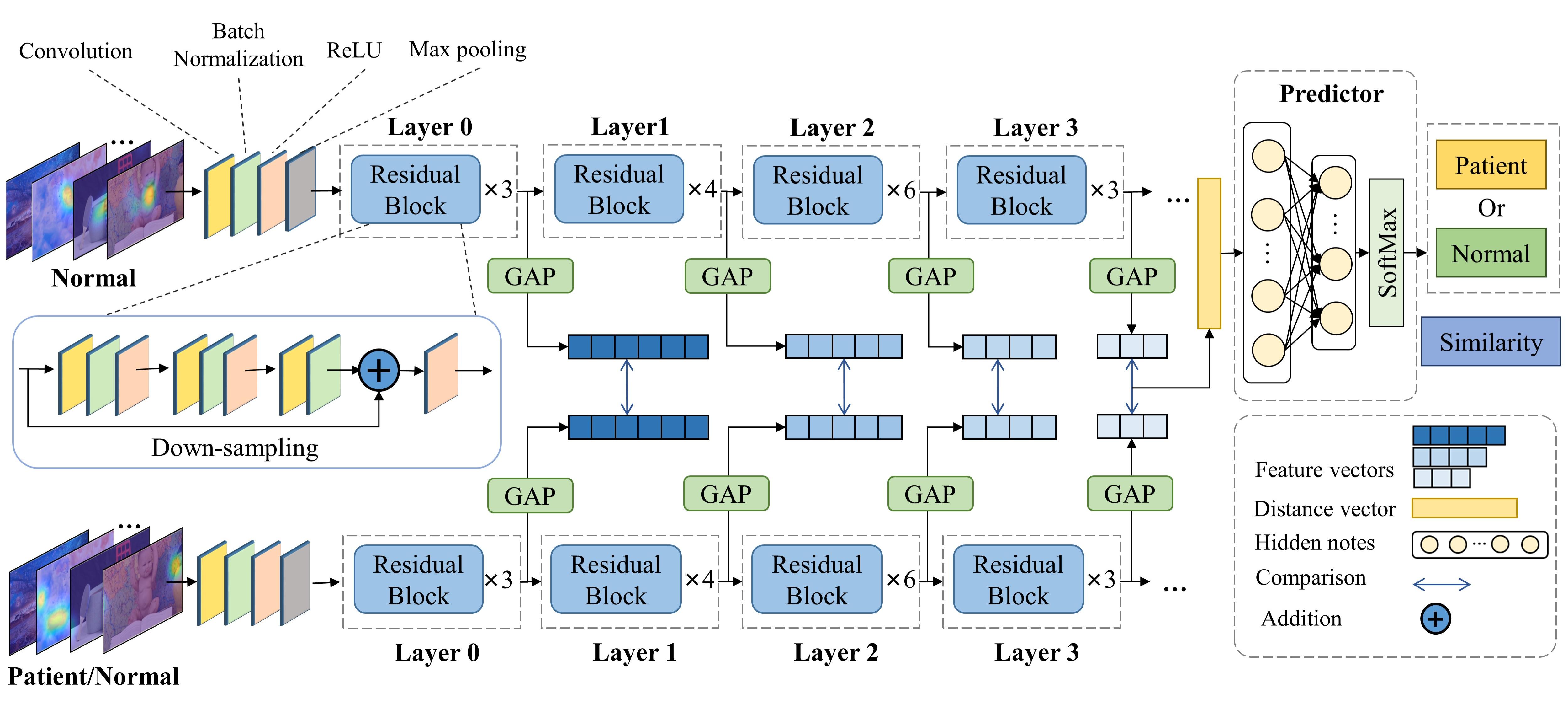}
\caption{An illustration of MC-CNN structure, consisting of two main components: feature extractor and predictor.}
\end{figure*}

\subsubsection{Multi-layered feature representation and integration}
The feature extractor is designed to explore multi-layered feature representations of heatmap combinations from shallow to deep layers. Shallower layers extract basic features such as intensity, color, and shapes, while deeper layers learn semantically stronger feature representations. All the features are then fused to represent the visual attention information of a heatmap combination. 

We take an example of a heatmap combination $\textbf{H}=[\textbf{X}_0,\textbf{X}_1]$, where matrices $\textbf{X}_0,\textbf{X}_1\in{R^{C\times H\times W}}$. $\textbf{X}_0$ and $\textbf{X}_1$ encode the stacked heatmaps of the unknown viewer (AD patient or normal) and the normal in a combination separately. $C$ represents the color channel dimension of the heatmap matrices, while $H$ and $W$ separately represent the height and width dimensions of the heatmaps. The function of feature-extracting layers can be defined as,
\begin{equation}
    \{\textbf{F}_i^0,\textbf{F}_i^1,\dots,\textbf{F}_i^j\}=f_{ext}(\textbf{X}_i),i\in{\{0,1\}},
\end{equation} where $f_{ext}(\cdot)$ is the function of the feature extractor, $\textbf{F}_i^j$ is the feature map obtained from the $i$th viewer in the combination on the $j$th layer.

GAP layers are connected at the end of each layer of feature extractors to encode the feature maps into feature vectors,  
\begin{equation}
    \textbf{v}_i^j=f_{GAP}(\textbf{F}_i^j),i\in{\{0,1\}},
\end{equation} where $f_{GAP}(\cdot)$ is the function of GAP layer and $\textbf{v}_i^j$ encodes the feature vector of the $i$th viewer in the combination on the $j$th layer. The distance of feature vectors from two viewers in a combination is calculated into a distance vector $\textbf{d}$, which represents the similarity between the heatmaps of two viewers, 
\begin{equation}
    \textbf{d}=[d^{0},d^{1},\dots,d^{j},\dots],
\end{equation}
\begin{equation}
    d^{j}=f_{dis}(\textbf{v}_0^j,\textbf{v}_1^j),
\end{equation} where $f_{dis}(\cdot)$ is the function to calculate the relevance of two feature vectors and $d^j$ encodes the relation of two heatmaps in the combination on the $j$th layer. In distance vector $\textbf{d}$, the relevances are arranged by the depth of convolution layers. Vector $\textbf{d}$ is fed to the predictor for feature integration.

\subsubsection{Classification for diagnosis}
The distance vector $\textbf{d}$ obtained from the feature extractor is then fed into an MLP that performs dimensionality reduction to learn the similarity information. A Softmax regression layer is connected at the end of MLP for binary classification. The 2-layered MLP consists of two perceptions that compress the relation vector $\textbf{d}$ into a 2-dimensional output vector $\textbf{s}$,
\begin{equation}
    \textbf{s}=[s_0,s_1]=\textbf{W}_2ReLU(\textbf{W}_1\textbf{d}+\textbf{b}_1)+\textbf{b}_2,
\end{equation}
where $\textbf{W}_1$, $\textbf{W}_2$ and $\textbf{b}_1$, $\textbf{b}_2$ represent the weight matrices and bias vectors of the first and second perception layers. $ReLU(\cdot)$is the activation function of the first perception layer. The vector $\textbf{s}$ is sent to the Softmax logistic regression layer. The output of Softmax layer is defined as $\hat{\textbf{l}}$:
\begin{equation}
    \hat{\textbf{l}}=[\hat{l_0},\hat{l_1}]=Softmax(\textbf{s})
\end{equation}
\begin{equation}
\hat{l_i}=P(Y=i|\textbf{s})=\frac{e^{s_i}}{e^{s_0}+e^{s_1}},i\in{\{0,1\}}
\end{equation}where $P(Y=i|\textbf{s})$ is the probability that the combination is classified as $Y$ in the case that the input vector is $\textbf{s}$. $Y=0$ means that the unknown viewer in the input combination is an AD patient, while $Y=1$ means he/she is a normal.

Since the higher $\hat{l_1}$ is, the greater probability that the unknown viewer is a normal, we define $\hat{l_1}$ as the similarity between the two heatmaps of the input combination $\textbf{H}$. The label of a combination is set as $l\in{\{0,1\}}$ which also serves as the targeted numerical value that similarity $\hat{l_1}$ needs to fit with. The fitting loss function is calculated by: 

\begin{equation}
    loss=l\log{\hat{l_1}}+(1-l)\log{⁡
    (1-\hat{l_1})}
\end{equation}

In the convergence procedure, the parameters of MC-CNN are updated by backward gradient descent, which propagates the gradient information of each parameter through the entire MC-CNN and optimizes all the parameters according to the fitting loss. Parameters could be updated in each training ergodic of all items in the training dataset:

\begin{equation}
    \mathbb{W}=\{w_0,w_1,\dots,w_i,\dots\}=\mathbb{W}-\alpha\frac{\partial{loss}}{\partial{\mathbb{W}}},
\end{equation}where refers $\mathbb{W}$ to all parameters that can be updated and $\alpha$ decides the rate of parameters updating, namely learning rate. With enough epoches, all parameters will be updated until the loss function reaches its minimum value, which indices that the proposed model is at its optimum state of best-fit the predictions to labels.

\subsection{Classification models for comparison}
 We compared our proposed scheme with several existing state-of-art classification methods based on the original/augmented heatmap dataset as shown in Fig. 6. For the original heatmap dataset, we employed several traditional convolutional neural networks for classification. Feature fusion and classification are carried out by the inner structure of each network. As for the augmented heatmap dataset, we adopted the convolutional feature-extrating module inside the networks to form the feature extractor. And then, we compared the feature representations on different layers to get distances of the input heatmap combinations, just as MC-CNN. The final classification was also finished by the same predictor in MC-CNN. The networks used for comparison included AlexNet, GoogLeNet, VGG11, and ResNet34:
\begin{itemize}
\item \textit{AlexNet.} AlexNet attempts to capture new and unusual features of the input images for classification using an architecture of 5 convolutional and 3 fully connected layers, which adopts several effective techniques such as Rectified Linear Units (ReLUs), local response normalization, and dropout for reducing training time and preventing overfitting;
\item \textit{GoogLeNet.} GoogLeNet is a deep convolutional neural network architecture to address the general image classification and detection problem, in which an efficient deep neural network architecture for computer vision codenamed Inception, is utilized to enforce the agreement with limited computational resources;
\item \textit{VGG11.} VGG11 is a deep hierarchical configuration of stacked convolutional layers followed by three fully connected layers, which has been developed to the depth of convolutional architecture via very small convolution filters in all layers;
\item \textit{ResNet34.} ResNet34 contains a stack of residual network architectures that have addressed the common degeneration problem of deep convolutional neural networks. In ResNet34 a deep residual learning framework is introduced to ease the training with considerably increased depth and gained accuracy;
\end{itemize}

\begin{figure}[htbp]  
\begin{center}  
\includegraphics[width=3.5in]{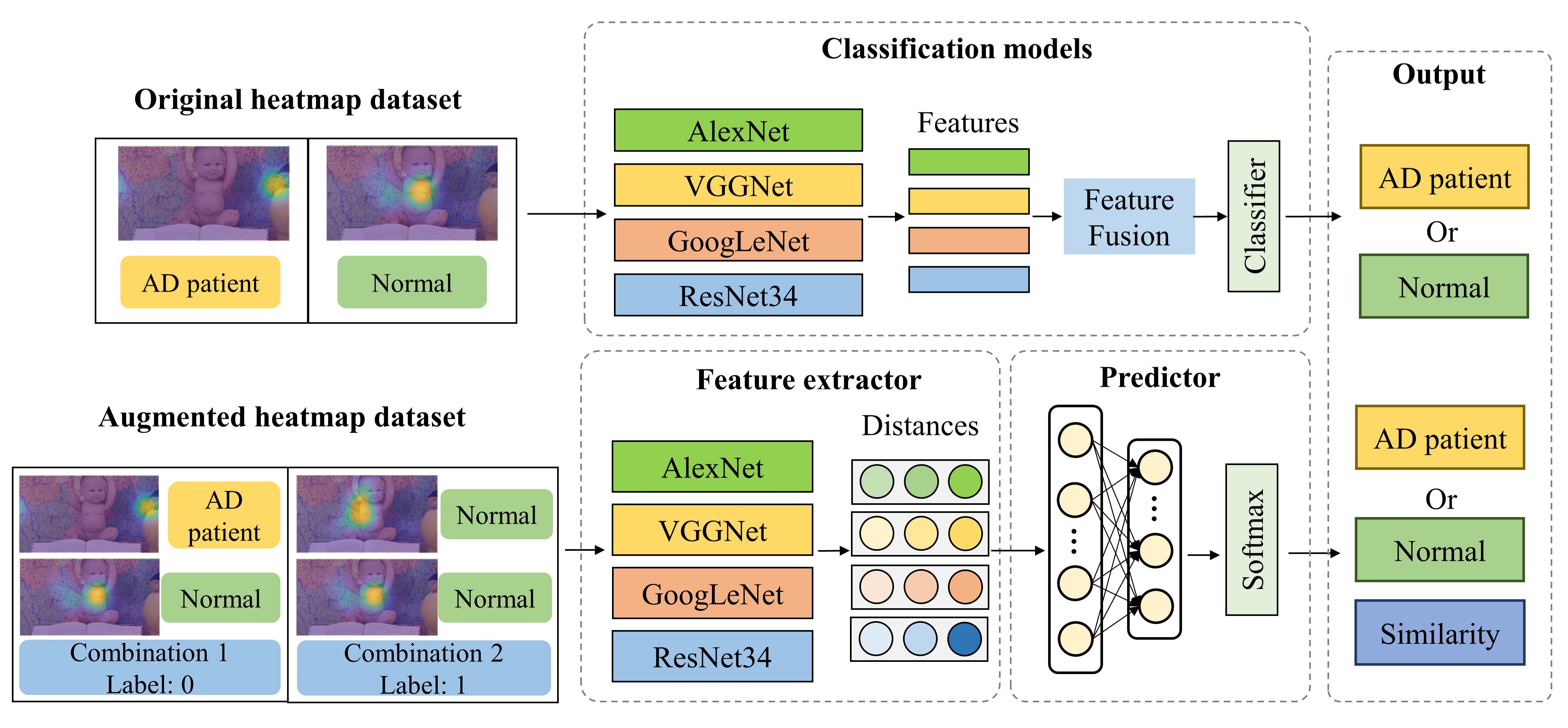}
\caption{The main framework of the networks separately based on the original heatmap dataset and the augmented heatmap dataset for comparisons.}  
\end{center}  
\end{figure}

\subsection{Evaluation metrics}
In this experiment, we introduced four assesment criteria, i.e., accuracy, recall, precision, and F1-score, to evaluate the performance of MC-CNN. The metrics are calculated according to Eq. (12)-(15) below:  
\begin{equation}
Accuracy=\frac{TP + TN}{TP + TN + FP + FN}
\end{equation}
\begin{equation}
Recall=\frac{TP}{TN + FP}
\end{equation}
\begin{equation}
Precision=\frac{TP}{TP + FP}
\end{equation}
\begin{equation}
F1\mbox{-}score=2\times{\frac{Precision\times{Recall}}{Precision + Recall}}
\end{equation}where TP, FP, TN, and FN are the calculated true positives, false positives, true negatives, and false negatives separately. In this paper, we took AD patients as the positive samples and normals as the negative samples. 

\section{Results}
\subsection{Convergence analysis}
The value of loss function and accuracy in the training phase and validation phase are logged by diagrams, presenting the process of model convergence. Training and validation loss will decent gradually with the number of epoches, in the meanwhile, training and validation accuracy will increase gradually, and then stabilize around a certain value as illustrated in Fig. 7.

\begin{figure}[htbp]
\centering
\includegraphics[width=3in]{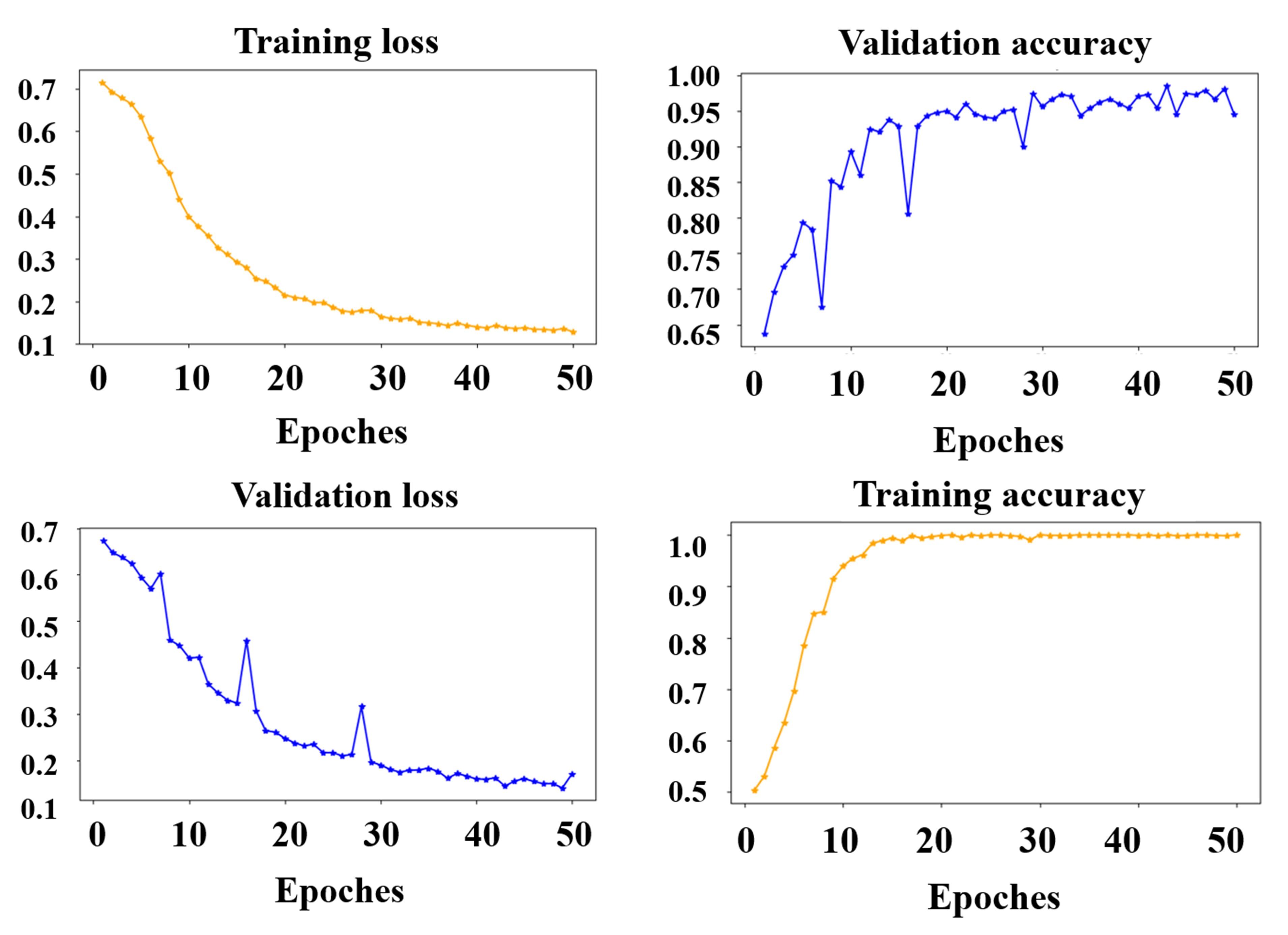}
\caption{The procedures of model convergence and validation.}
\end{figure}

\subsection{Adjustmentations of parameters}
\subsubsection{Learning Rate}
The learning rate operates as the speed of gradient descent, which the performance of model is sensitive to. We tried different learning rate values ranging from a magnitude of $10^{-5}$ to $10^{-2}$ in the network. We evaluated the accuracy under different learning rates on several heatmaps for test, and the result is shown in Fig. 8. Considering the characteristic of the learning rate value, the statistical data was logged as logarithmic increasing. From Fig. 8, the optimistic result distributes around the value $10^{-3}$ . This gives the appropriate range of learning rate to subtly adjust.

\begin{figure}[htbp]
\centering
\includegraphics[width=3in]{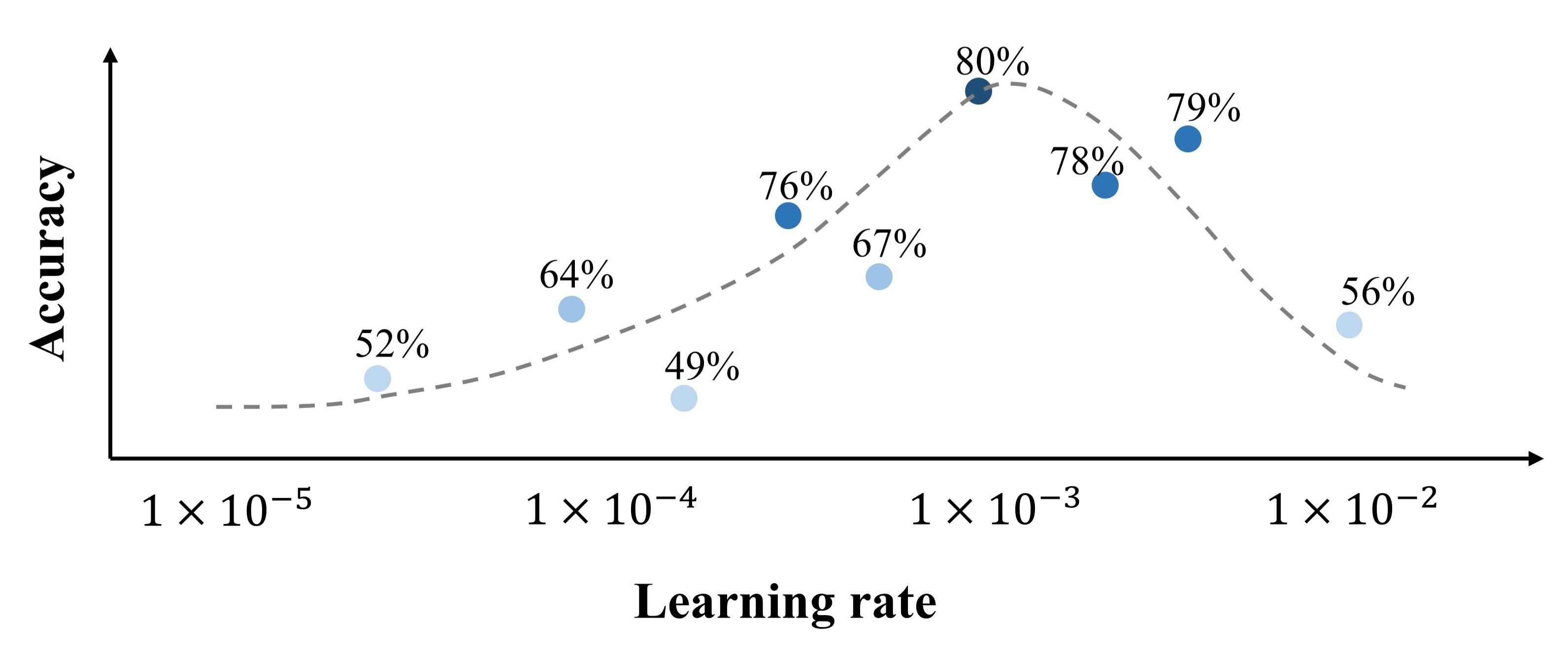}
\caption{Test accuracy with different learning rate values.}
\end{figure}

\subsubsection{Number of convolutional feature-extracting layers}
As the convolutional feature extracting layer goes deeper, the feature map of which backward represents information of different semantic strengths. The opportune number of convolution layers can bring the most accurate classification results. In this experiment, the feature extractor of MC-CNN generates feature representations of five layers, which are coded from 0 to 4. The performance of different combinations of feature-extracting layers is shown in TABLE II, from which we can see that the more feature representations are integrated, the better performance can MC-CNN gets. Therefore, the MC-CNN model is finally set to integrate feature representations from 5 feature extracting layers. 

\begin{table}[htbp]
\caption{Performances with different combinations of feature extracting layers.}
\centering
\begin{tabular}{ccccc}
\toprule
Metrics &Layer01 & Layer012 & Layer0123 & Layer01234\\
\midrule
Accuracy & 0.60$\pm$0.02 & 0.68$\pm$0.01 & 0.79$\pm$0.01 & \textbf{0.83}$\pm$\textbf{0.01} \\
Recall & 0.58$\pm$0.06 & 0.60$\pm$0.06 & 0.57$\pm$0.03 & \textbf{0.74}$\pm$\textbf{0.01}\\
Precision & 0.75$\pm$0.07 & 0.79$\pm$0.04 & 0.99$\pm$0.01 & \textbf{0.90}$\pm$\textbf{0.01}\\
F1-Score & 0.59$\pm$0.01 & 0.64$\pm$0.01 & 0.72$\pm$0.02 & \textbf{0.81}$\pm$\textbf{0.01}\\
\bottomrule
\end{tabular}
\end{table}

To ensure the robustness of the model, we took three-fold cross-validation to check the performance of MC-CNN. The accuracy of the convergence, validation and test phases of the three folds is presented in table III.

\begin{table}[htbp]
\caption{Training accuracy, validation accuracy, and
 test accuracy in the three folds.}
 \centering
\begin{tabular}
{cccc}
\toprule
Folds & Training accuracy & Validation accuracy & Test accuracy\\
\midrule
Fold 1 & 0.99 & 0.72 & 0.86 \\
Fold 2 & 0.99 & 0.79 & 0.81 \\
Fold 3 & 0.97 & 0.75 & 0.81 \\
\textbf{Average value} & \textbf{0.99$\pm$0.01} & \textbf{0.753$\pm$0.01} & \textbf{0.83$\pm$0.01}\\
\bottomrule
 \end{tabular}
\end{table}

\subsection{Classification results of different models}
We tested the performance of networks for comparison (see Section II-E) on the same test dataset by applying three-fold cross-validation. Table IV reports the classification performances of our proposed method and other classification models separately based on the original/augmented heatmap dataset. Fig. 9 shows the classification results of 16 participants (8 AD patients ad 8 normals) that are randomly selected from the test dataset calculated by three different models. The 8 AD patients and 8 normals are numbered sequentially from 1 to 8 in the abscissa axis, and are represented by the spots of two shapes(as shown in Fig. 9). The ordinate value of each individual represents the similarity value (for augmented dataset) or the output value before the classifier (for the original dataset). 

\begin{figure}[htbp]  
\begin{center}  
\includegraphics
[width=3.5in]{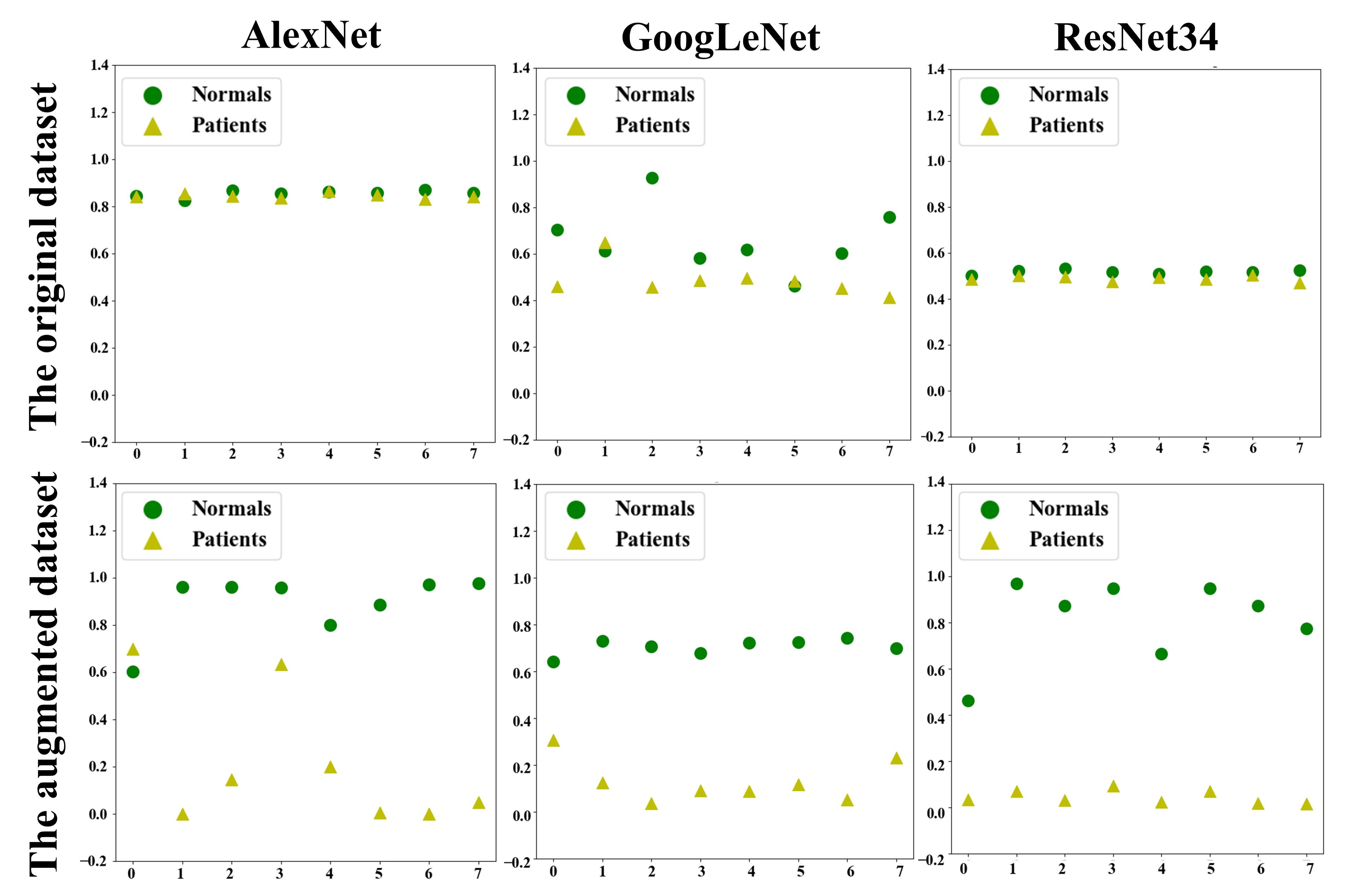}
\caption{The classification performances of three comparison models of 16 subjects (8 patients and 8 normals) respectively based on the original/augmented dataset.}
\end{center}  
\end{figure}

\begin{table}[htbp]
	\centering
	\caption{Performance comparison between our proposed method and several state-of-art classification networks.}
	\begin{tabular}{cccc}
		\toprule  
		Model & Metrics & Original dataset& Augmented dataset \\ 
		\midrule  
        \multirow{5}{*}{AlexNet}
		& Accuracy & 0.63$\pm$0.01 & 0.68$\pm$0.01 \\
        & Recall & 0.31$\pm$0.03 & 0.48$\pm$0.04 \\
        & Precision & 0.83$\pm$0.08 & 0.80$\pm$0.03 \\
        & F1-score & 0.26$\pm$0.02 & 0.59$\pm$0.04 \\
        \midrule  
        \multirow{5}{*}{VGG11}
		& Accuracy & 0.74$\pm$0.05 & 0.79$\pm$0.01 \\
        & Recall & 0.48$\pm$0.20 & 0.66$\pm$0.03 \\
        & Precision & 0.83$\pm$0.08 & 0.90$\pm$0.01 \\
        & F1-score & 0.78$\pm$0.02 & 0.75$\pm$0.01 \\
        \midrule  
        \multirow{5}{*}{GoogLeNet}
		& Accuracy & 0.74$\pm$0.01 & 0.72$\pm$0.01 \\
        & Recall & 0.60$\pm$0.01 & 0.61$\pm$0.01 \\
        & Precision & 0.86$\pm$0.02 & 0.81$\pm$0.03 \\
        & F1-score & 0.68$\pm$0.01 & 0.69$\pm$0.01 \\
        \midrule  
        \multirow{5}{*}{ResNet34}
		& Accuracy & 0.72$\pm$0.01 & 0.81$\pm$0.01 \\
        & Recall & 0.60$\pm$0.11 & 0.70$\pm$0.01 \\
        & Precision & 0.83$\pm$0.02 & 0.91$\pm$0.03 \\
        & F1-score & 0.65$\pm$0.03 & 0.79$\pm$0.01 \\
        \midrule  
        \multirow{5}{*}{\textbf{MC-CNN(ours)}}
		& Accuracy & \multirow{4}{*}{$-$} 
        & \textbf{0.83$\pm$0.01}\\
        & Recall &  & \textbf{0.74$\pm$0.01}\\
        & Precision &  & \textbf{0.90$\pm$0.01}\\
        & F1-score &  & \textbf{0.81$\pm$0.01}\\
		\bottomrule  
	\end{tabular}
\end{table}

From Fig. 9, it can be seen that the classification models perform better on the augmented dataset compared to the original dataset, which is manifested as the farther distances of two kinds of spots. It is worth noting that AlexNet (with a simple convolutional feature extractor) performs the worst, while ResNet34 (with residual convolutional feature extracting structure) shows the best classification result. The following observations can be made from TABLE IV: 1) Our proposed MC-CNN performs the best among all the comparative models. 2) AlexNet, as network with the fewest convolutional layers, performs the worst on the original dataset, indicating that feature representations on shallow feature extracting layers only are insufficient to accurately separate between AD patients and normals. 3) After employing the dataset augmentation, the AlexNet, VGG11, and ResNet34 provide a significant improvement in classification accuracy. 4) Although GoogLeNet performs no improvements in accuracy, the recall value has improved on the augmented dataset, which is also an important indicator in disease detection. 5) The AD patients and normals can get better discrimination on the augmented dataset, which further verifies the effectiveness of data augmentation and the proposed MC-CNN model. 

\subsection{Ablation experiment for the modules in MC-CNN}
To more comprehensively evaluate the performance of MC-CNN, we designed an ablation experiment to verify the importance of each module of the MC-CNN. Since the MC-CNN extracts multi-layered feature representations using multiple residual blocks and GAP (as shown in Fig. 5) based on the augmented dataset, we design the ablation experiment by replacing them with other modules. For the augmented dataset, we replace it with the original dataset. As for the multiple representations, we change them to a single representation of the last residual block. Moreover, we change the GAP layers into fully-connected layers. Thus, there is a total of four models for comparison: Model1 (model based on the original dataset), Model2 (model with feature representations from a single residual layer), Model3 (model with fully-connected layers rather than GAP layers), and Model4 (the proposed MC-CNN model). The designation of the ablation experiment is presented in Fig. 10, and TABLE V indicates the performances of the four models.

\begin{figure}[htbp]  
\begin{center}  
\includegraphics[width=3.5in]{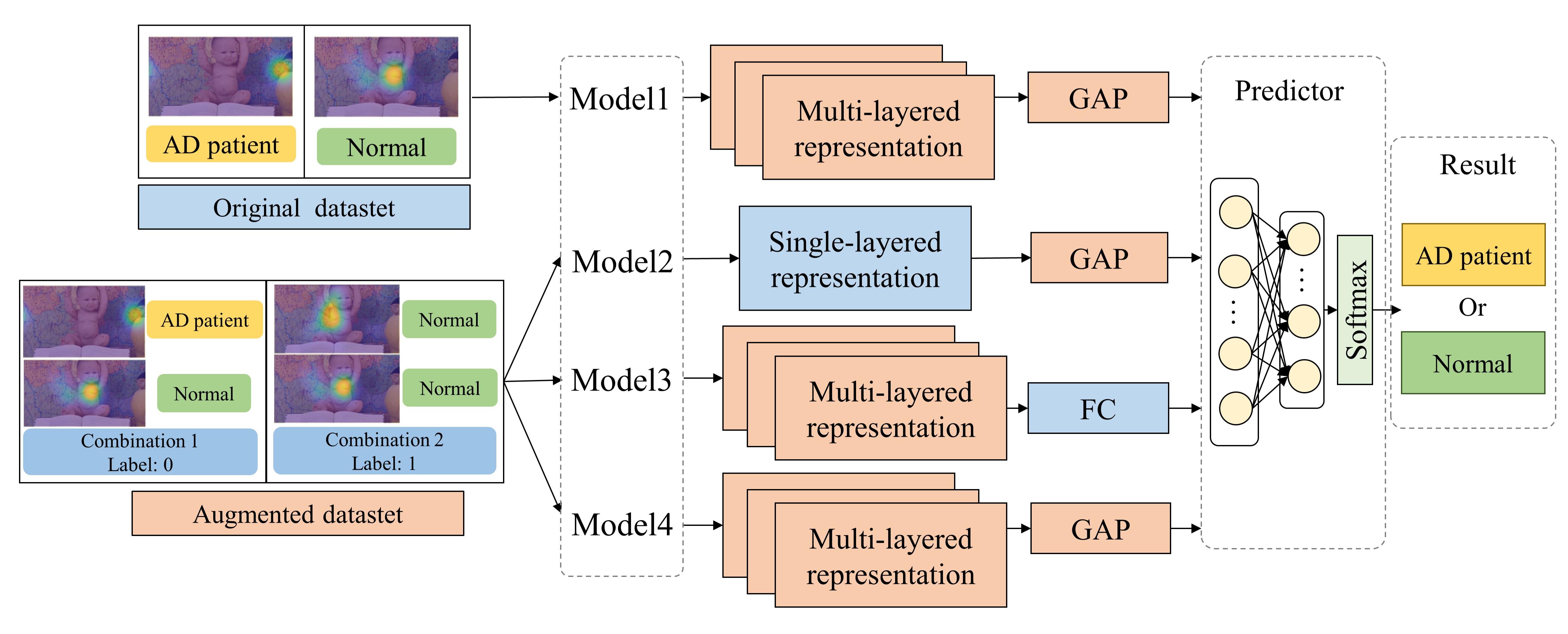}
\caption{The flowchart of four different models for comparison in the ablation experiment.}  
\end{center}  
\end{figure}

\begin{table}[htbp]
\caption{Ablation experiment results for the proposed MC-CNN.}
\centering
\begin{tabular}{ccccc}
\toprule
Models & Accuracy & Recall & Precision & F1-score\\
\midrule
Model1 & 0.72$\pm$0.01 & 0.88$\pm$0.01 & 0.66$\pm$0.01 & 0.79$\pm$0.01 \\
Model2 & 0.81$\pm$0.01 & 0.79$\pm$0.07 & 0.89$\pm$0.04 & 0.80$\pm$0.02\\
Model3 & 0.74$\pm$0.01 & 0.69$\pm$0.03 & 0.79$\pm$0.02 & 0.73$\pm$0.01 \\
\textbf{Model4(ours)} & \textbf{0.83$\pm$0.01} & \textbf{0.74$\pm$0.01} & \textbf{0.91$\pm$0.01} & \textbf{0.81$\pm$0.01}\\
\bottomrule
\end{tabular}
\end{table}

As shown in TABLE V, four models show different performances for classifying AD patients and normals. Among them, Model4 shows the best performance, achieving 0.827$\pm$0.001 accuracy. Model1, which is based on the original dataset rather than the augmented dataset, shows the worst classification accuracy of 0.723 $\pm$ 0.014. By replacing the multi-layered feature extractor and the GAP layers in turn, the classification accuracy of the two models (Model2 and Model3) is decreased by 0.017 and 0.089, respectively. 

The reasons for the differences in the performance of these models can be attributed to the following points: First, for the multi-layered feature extractor, by comparing the multiple feature representations of heatmaps, the model learns the shallower representations as supplement information, which can provide the classifier with more comprehensive features, thus achieves higher accuracy. Second, using the GAP layers rather than traditional fully-connected layers for feature integration introduces no parameter to optimize, thus the overfitting problem is alleviated. Most importantly, for the augmented dataset, the size of the training dataset is significantly enlarged which allows the proposed model to more sufficiently learn the similarity features between the heatmaps in combinations. However, the original heatmap dataset is relatively not large enough to support the model with sophisticated structures. 

\section{Conclusion and Future Works}
In this paper, we have proposed a novel deep-learning-based classification framework to alleviate the low efficiency, invasiveness, and manpower requirements in the early diagnosis of AD. By taking advantage of a 3D comprehensive visual task and multi-layered convolutional neural network, we efficiently integrated the characteristic eye-tracking features extracted from AD patients and normals into a distance representation and achieved enhanced robust classification results for analyzing visual attention deficits in AD patients. We also designed a dataset augmentation method to enlarge the size of training dataset for more sufficient model learning. Experimental results on the recruited subjects demonstrated that the performance of our proposed scheme obtained superior performance over the state-of-art methods.

In the future, we will build a larger and more comprehensive eye-movement dataset by including AD patients with different degrees of cognitive impairments. Besides, different models will be designed to analyze more eye-movement features with the updated dataset. 

\section*{Acknowledgement}
The authors would like to thank all of the study participants and their families that participated in the research on Alzheimer's disease here.
\bibliographystyle{IEEEtran}
\bibliography{Ref}

\vfill
\end{sloppypar}
\end{document}